\documentclass[fleqn,usenatbib]{mnras}

\usepackage{amssymb}	

\usepackage{newtxtext,newtxmath}

\usepackage[T1]{fontenc}
\usepackage{ae,aecompl}

\usepackage{multirow}
\usepackage{graphicx}	
\usepackage{amsmath}	
\usepackage{hyperref}

\title[GW catalogue of TDEs]{Gravitational waves from tidal disruption events: an open and comprehensive catalogue}

\author[Toscani M. et al.]{
Martina Toscani,$^{1,2}$\thanks{E-mail: martina.toscani@unimi.it}
Giuseppe Lodato$^{1}$,
Daniel J. Price$^{3}$ and
David Liptai$^{3}$
\\
$^{1}$ Dipartimento di Fisica, Universit\`a Degli Studi di Milano, Via Celoria, 16, Milano, 20133, Italy\\
$^{2}$ Laboratoire des 2 Infinis - Toulouse (L2IT-IN2P3), Universit\'e de Toulouse, CNRS, UPS, F-31062 Toulouse Cedex 9, France\\
$^{3}$ School of Physics and Astronomy, Monash University, Clayton Vic 3800, Australia\\
}

\date{Accepted XXX. Received YYY; in original form ZZZ}

\pubyear{2021}

\begin{document}
\label{firstpage}
\pagerange{\pageref{firstpage}--\pageref{lastpage}}
\maketitle

\begin{abstract}
We present an online, open and comprehensive template library of gravitational waveforms produced during the tidal disruptions of stars by massive black holes, spanning a broad space of parameters. We build this library thanks to a new feature that we implement in \textcolor{black}{the general relativistic version of \textsc{phantom}}, a smoothed particle hydrodynamics code for three dimensional simulations in general relativity. We first perform a series of numerical tests to show that the gravitational wave (GW) signal obtained is in excellent agreement with the one expected from theory. This benchmark is done for well studied scenarios (such as binary stellar systems). We then apply our code to calculate the GW signals from tidal disruption events (TDEs), finding that our results are consistent with the theoretical estimates obtained in previous studies for selected parameters. We illustrate interesting results from the catalogue, where we stress how the gravitational signal is affected by variations of some parameters (like black hole spin, stellar orbital eccentricity and inclination). The full catalogue is available online. It is intended to be a living catalogue.
\end{abstract}

\begin{keywords}
gravitational waves -- hydrodynamics -- methods: numerical -- black hole physics -- accretion, accretion discs
\end{keywords}



\section{Introduction}

Simultaneous detection of the neutron star merger in 2017, both in gravitational waves (GWs, \citealt{Abbott:17aa}) and electromagnetic (EM) radiation \citep{Goldstein:17aa}, has paved the way to an era of multi-messenger astronomy. In the near future, a new generation of interferometers, including LISA \citep{Amaro-Seoane:17aa}, TianQin \citep{Luo:16aa} and DECIGO \citep{Sato:17aa}, will be launched in space to reveal the low-frequency ($10^{-4}-10^{-2}\,\text{Hz}$) GW sky for the first time. At the same time, powerful telescopes, such as Athena \citep{Barcon:12aa}, Lynx \citep{Gaskin:18aa} and LSST \citep{LSST:09aa}, will search the sky looking for EM counterparts of these gravitational sources.\\
\indent Among the systems we expect to detect in both domains are stars tidally disrupted by black holes (BHs). Tidal disruption events (TDEs; \citealt{Rees:88aa}, \citealt{Phinney:89aa}) have been extensively investigated over the last decades both analytically (see the review by \citealt{Rossi:20aa} and references therein) and numerically (see the review by \citealt{Lodato:20aa} and references therein). To date, around 50 events of this type have been seen, in different bands of the EM spectrum (optical: \citealt{vanvelzen:2020}, X-ray: \citealt{Saxton:20}, radio: \citealt{Alexander:20}). They are characterized by luminous flares, often super-Eddington, that arise when the stellar debris falls back to the pericenter, typically several days after the star is torn apart by the BH tides.\\
\indent As for the destruction itself, so far we have not been able to observe it since there is little EM production during that phase. Yet, a star undergoing the disruption emits a GW burst. For a standard scenario of a Sun-like star disrupted by a $10^{6}\,\text{M}_{\odot}$ BH, we expect the GW signal to be $\approx 10^{-22}$, with a frequency $\approx 10^{-4}\,\text{Hz}$ (see, e.g., \citealt{Kobayashi_2004}, \citealt{Toscani:19aa}). Thus, the frequency is in the right interval for space-based interferometers. Still, the signal is not very strong and for this reason we do not expect LISA to see many events of this type (between zero and a few tens during the whole mission, a quantity \textit{strongly} dependent on the noise sensitivity of the instruments; see \citealt{Pfister:21aa}). Contrarily, the post-LISA generation of interferometers (e.g. DECIGO) will be more sensitive and thousands of TDEs will be seen in GWs \citep{Pfister:21aa}. Hence it is certain that in the future we will have the proper instruments to see the GW emission from TDEs.\\
\indent For this reason, it is important to have a code that not only can predict the GW emission from a source, but also manages to reproduce the fluid-dynamics accurately. Smoothed particle hydrodynamics (SPH) (\citealt{Gingold:77aa}, \citealt{Lucy:77aa}, \citealt{Rosswog:09ab}, \citealt{Price:12aa}) is perfect for this goal. Yet, since we are also interested in gravitational emission, that typically involves compact objects in strong gravity, we need an SPH code with general relativistic features. A code that couples general relativity (GR) and hydrodynamics is \textsc{grphantom} \citep{Liptai:19aa}, the latest version of the \textsc{phantom} code \citep{Price:18aa}. \textsc{grphantom} works in any fixed metric (Minkowski, Schwarzschild and Kerr in Cartesian Boyer-Lindquist coordinates), with a proper treatment of orbital dynamics and with the ability to capture relativistic shocks. It is open source and easily available for everyone.\\
\indent Here we describe our implementation of a module\footnote{It will be soon freely available within \textsc{grphantom}.} to compute the GW emission in \textsc{grphantom} and use it to create a library of TDE waveforms. In Section \ref{sec:2} we describe the method, while in section \ref{sec:3} we present standardised test. In Section \ref{sec:4} we discuss the basics of TDE physics and we apply the tool to derive the GW signal produced by the standard scenario of this kind of event. Furthermore, we also show how the GW catalogue has been built (available at this website \url{https://gwcataloguetdes.fisica.unimi.it}). Finally, we discuss some interesting results in Section \ref{sec:5} and we draw our conclusions in Section \ref{sec:6}.
\begin{figure}
    \centering
    \includegraphics[width=0.4\textwidth]{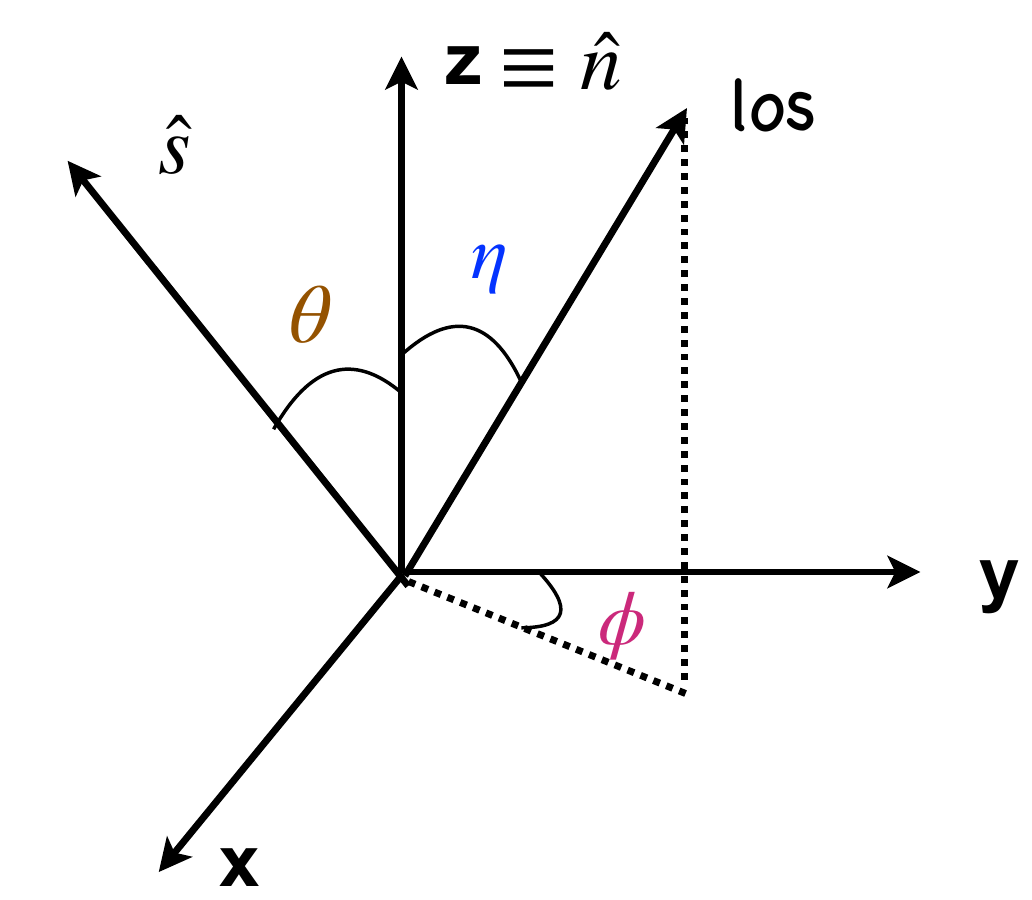}
    \caption{Relations between the $(\pmb{x},\pmb{y},\pmb{z})$ frame and the angles $\eta, \phi, \theta$. For our TDEs, the stellar orbit lies in the $\pmb{x}-\pmb{y}$ plane, the normal to the orbit $\hat{n}$ is along the $\pmb {z}$ direction. $\eta$ is the angle between the line of sight (los) and the normal to the orbit, $\phi$ is the angle among the projection of the los on the $\pmb{x}-\pmb{y}$ plane and the pericenter of the star. $\theta$ is the angle between the spin of the BH and the normal to the orbit (see Section \ref{sec:4}). }
    \label{fig:angles}
\end{figure}

\section{Method}
\label{sec:2}
Assuming the weak field approximation and the Lorenz gauge, the gravitational wave emission is described by \textcolor{black}{(see, e.g., \citealt{Einstein:18aa}, \citealt{Buonanno:07aa} and \citealt{Toscani:19aa})}
\begin{align}
    h^{\rm TT}_{ij}(t,\pmb x)=\frac{2G}{dc^4}\Lambda_{ij,kl}\ddot{M}^{kl}(t,\pmb x),
    \label{eq:strain}
\end{align}
where $h^{\rm TT}_{ij}$ is a small perturbation of the background metric {(for the TDEs presented here, we use the Kerr metric)}, $G$ is the gravitational constant, $c$ is the speed of light, $d$ is the distance of the source with respect to the observer, $\Lambda_{ij,kl}$ is the TT operator\footnote{The superscript $\rm TT$ means that the Transverse Traceless gauge holds.} and $\ddot{M}^{kl}$ is the second time derivative of the \textcolor{black}{moment} of inertia of the system
\begin{align}
    M^{kl}=\int{\textcolor{black}{\text{d}^3 x} \rho(t,\pmb x)x^{k}x^{l}}.
    \label{eq:intertiamom}
\end{align}
The index $k$ and $l$ run from 1 to 3 and stand for the spatial coordinates.\\
\indent Only two components of $h^{\rm TT}_{ij}$ are independent. These components are the \textit{cross polarization} $h_\times$, and the \textit{plus polarization}, $h_{+}$. 
If we consider the GW propagating along a generic direction $\pmb k$, the quadrupole radiation will depend on two angles (see, e.g., \citealt{Maggiore:07aa}). Here, we call these angles $\eta$ and $\phi$. We define $\eta$ as the angle between the line of sight and the $\pmb{z}$ direction, that we will always consider to be perpendicular to the stellar orbit in our TDE calculations\footnote{This direction can be different from the one of spin axis. For further details see the final part of Section \ref{sec:4}.}. Instead $\phi$ is the angle between the projection of the line of sight on the orbital plane and the $\pmb{y}$ axis, along which we will always assume the stellar pericenter to lie. These angles are illustrated in Figure \ref{fig:angles}, where we show also a third angle, $\theta$, that we define later in the paper (cf. Section \ref{sec:4}). \\
\indent Note that, in the rest of the paper, we will call \textit{face-on} polarization amplitudes
\begin{align}
   h_+&=\frac{G}{\textcolor{black}{d}c^4}(\ddot M_{11}-\ddot M_{22}),\label{eq:h+}\\
    h_{\times}&=\frac{2G}{\textcolor{black}{d}c^4}\ddot M_{12}\label{eq:hx}, 
\end{align}
the specific case where the line of sight is along the $\pmb z$ direction and thus $\eta=0$ (cf. \citealt{Maggiore:07aa}).\\
\indent Since in \textsc{grphantom} we simulate astrophysical objects as an ensemble of SPH particles, the starting point to implement the calculation of the gravitational emission is to discretize formula \eqref{eq:intertiamom}. We proceed in the following way
\begin{align}
 M^{kl}=\int{\textcolor{black}{\text{d}^{3}x} \rho(t,\pmb x)x^{k}x^{l}} \rightarrow  \int{\text{d} m x^{k}x^{l}} \simeq \sum_{\rm b} m_{\rm b} x^{k}_{\rm b}x^{l}_{\rm b},
 \label{eq:disc_momin}
\end{align}
where in the last step we have introduced the index $\rm b$ that counts the number of particles. {Note that the positions are with respect to the origin of the coordinate system.} Starting from this formula, we can compute the second time derivative as 
\begin{align}
 \ddot{M}^{kl}=\sum_{b}m_{b}({a}_{l}x_{k}+2{v}_{l}{v}_{k}+x_{l} a_{k}), 
\end{align}
where $v$ and $a$ are the velocity and acceleration of the SPH particle. The code knows $v$ and $a$ at each timestep and consequently we can calculate \textit{on the fly} the relevant quantities for the GW emissions: $\ddot{M}^{kl}$ and the polarization amplitudes $h_+$ and $h_{\times}$ (for selected values of $\eta$ and $\phi$, see Section \ref{sec:4}). Each of these quantities is written in an output file at each time. {The acceleration $a$ is computed as the time derivative of the momentum per unit mass, that in GR is in general different from the time derivative of the velocity. Yet, we prefer to use this quantity, that is the acceleration term calculated in the code, since it is the best way to take into account GR effects. In any case, since in our calculation we do not consider ultra-close TDEs, these effects will not be particularly strong.}\\
\indent Note that we are neglecting the energy lost by the system via GWs (for further discussion on this, see Section \ref{sec:5}).

\begin{figure*}
    \centering
    \includegraphics[width=0.8\textwidth]{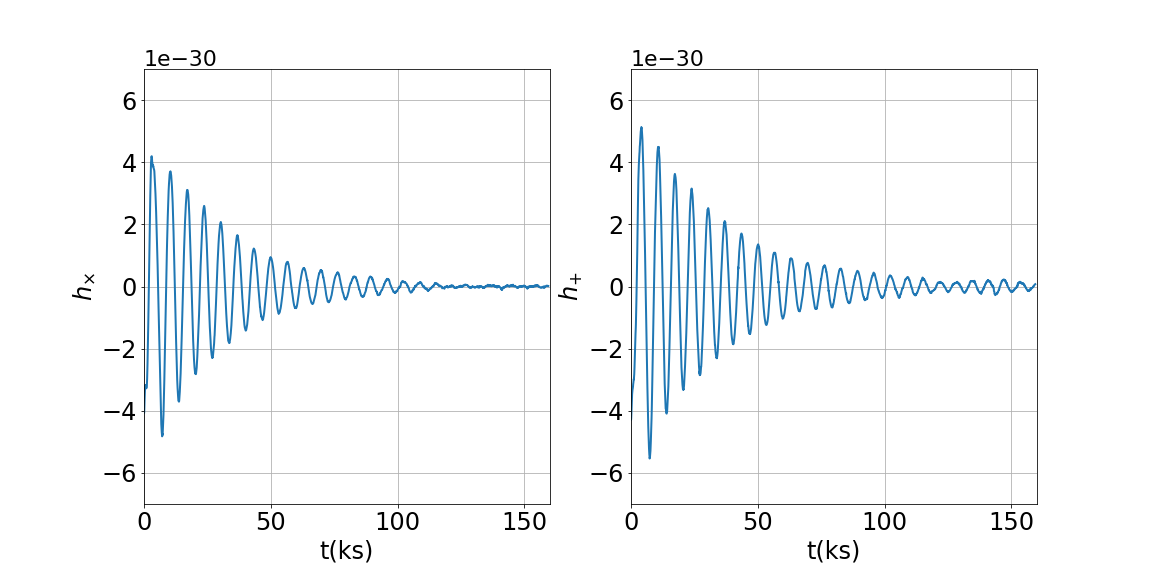}
\caption{GW polarization amplitudes by an isolated Sun-like star at a distance of $d=20\text{Mpc}$, plotted with respect to time (in kiloseconds). \textbf{Left}: $\times$ polarization. \textbf{Right}: $+$ polarization.} 
    \label{fig:star_strain}
\end{figure*}
\begin{figure*}
    \centering
   \includegraphics[width=0.8\textwidth]{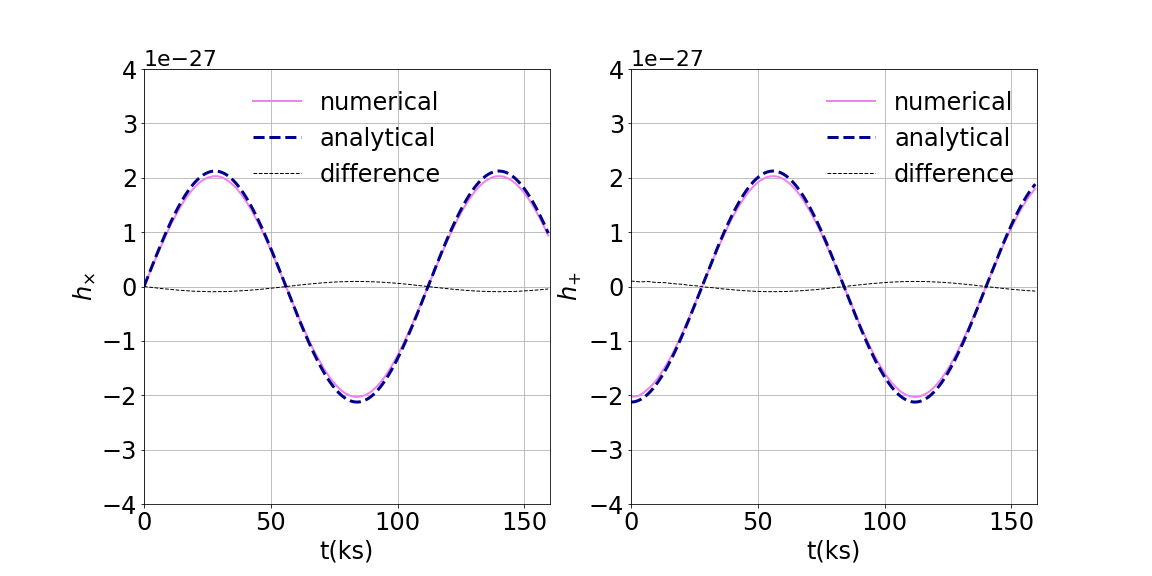}
\caption{GW polarization amplitudes by a binary system of two Sun-like stars, separated by $R=10 \text{R}_{\sun}$, at a distance of $d=20\text{Mpc}$ from us, plotted with respect to the time (in kiloseconds). \textbf{Left}: $\times$ polarization seen face-on. \textbf{Right}: $+$ polarization seen face-on. The pink line is the signal calculated during the simulation, while the blue dashed-line represents the analytical solution assuming point sources. \textcolor{black}{The black dashed line is the difference between the numerical and the analytical curves.}} 
    \label{fig:binary_strain}
\end{figure*}
\section{Tests}
\label{sec:3}
We perform a series of tests to check that our tool for the derivation of the quadrupole radiation works properly, in a similar \textcolor{black}{manner} to those in \citet{Loren:04aa}.\\
\indent First, we simulate an isolated star and we derive numerically its GW emission. Theoretically we expect the signal to be zero since the star is spherical and not moving. Yet this needs to be verified because SPH particles are not initially in equilibrium. We simulate a Sun-like star, assumed to be a polytropic sphere with index $\gamma=5/3$, at a distance of $20\text{Mpc}$ from us, using $\approx 10^4$ equal-mass particles. Our results are illustrated in Figure \ref{fig:star_strain}. On the left we show the $\times$ polarization, on the right the $+$ polarization. Amplitudes are plotted with respect to time, expressed in kiloseconds. \textcolor{black}{While the star is reaching equilibrium ($t<100$ ks), the GW emission produced by SPH particles decreases rapidly. After the equilibrium ($t>100$ ks), the emission has become at least one order of magnitude lower ($\sim 10^{-31}$). We can consider this residual emission due to SPH particles oscillations negligible, since much smaller than any signals produced by astrophysical sources of interest located at the same distance. For instance, in the case of a TDE of a Sun-like star disrupted by a $10^{6}\text{M}_{\odot}$ BH at 20 Mpc, this residual emission is 9 orders of magnitude lower than the GW burst associated to the disruption.}\\
\indent Second, we investigate the emission associated with a stellar binary, \textcolor{black}{assuming Newtonian gravity in Minkowski metric plus small perturbations}. We assume two stars of mass $M_1$ and $M_2$, separated by $R$, moving on a circular orbit with angular velocity $\omega$ at a distance $d$ from the observer. The GW emission is (see, e.g., \citealt{Maggiore:07aa})
\begin{align}
    h_{\rm \times}(t,d,\eta)&=\frac{4G\mu\omega^2R^2}{dc^4}\cos\eta\sin(2\omega t),\\
   h_{+}(t,d,\eta)&=\frac{4G\mu\omega^2R^2}{dc^4}\left( \frac{1+\cos^2\eta}{2}\right)\cos(2\omega t),
\end{align}
where $\mu=M_1M_2/(M_1+M_2)$ is the reduced mass of the system. For this scenario, we consider two Sun-like stars separated by $10\text{R}_{\sun}$, each simulated as a polytropic sphere with $\gamma=5/3$ and $10^4$ equal-mass particles. The system is located at a distance $d=20\,\text{Mpc}$. After relaxing the system, we let the simulation evolve for $\approx 1$ orbit. Figure \ref{fig:binary_strain} shows the emission with respect to time (in kiloseconds). In pink we show the signal derived with \textcolor{black}{\textsc{grphantom}} during the simulation, while the analytical solution is illustrated with the blue dashed-line. The layout of the panels is the same as in the previous figure. The plots show \textcolor{black}{agreement to within $\approx 4\%$} between the numerical and the analytical solution.


\section{Gravitational waves from tidal disruption events}
\label{sec:4}

\subsection{Physical properties of tidal disruption events}
\label{sec:theory}
The standard picture of a TDE (see, e.g., \citealt{Rees:88aa}, \citealt{Phinney:89aa}) considers a star with mass $M_*=m_*\text{M}_\odot$ and radius $R_*=r_*\text{R}_\odot$, on a Keplerian orbit around a massive BH $M_{\rm h}=m_{\rm h} \text{M}_\odot$ ($\text{M}_\odot$ and $\text{R}_\odot$ are the mass and the radius of the Sun respectively). The star is torn apart by BH tides if the stellar pericenter $r_{\rm p}$ falls in the range $r_{\rm s}\lesssim r_{\rm p} \lesssim r_{\rm t}$. Here $r_{\rm t}$ is the tidal radius, that is the distance where BH tides equal the stellar self-gravity,
\begin{align}
    r_{\rm t}\approx \left(\frac{M_{\rm h}}{M_*} \right)^{1/3}\approx 7\times 10^{10}\left(\frac{m_{\rm h}}{m_*}\right)^{1/3}r_*\,\text{cm},
\end{align}
and $r_{\rm s}$ is the BH Schwarzschild radius 
\begin{align}
    r_{s}=\frac{2GM_{\rm h}}{c^2}\approx 3m_{\rm h}\times 10^5\,\text{cm}.
\end{align}
The strength of the event is quantified by the penetration factor $\beta =r_{\rm t}/r_{\rm p}$. Following the previous considerations, $\beta$ spans the following interval 
\begin{align}
    1\lesssim \beta \lesssim \beta_{\rm max}=\frac{r_{\rm t}}{r_{\rm s}}=2\times 10^5 \, r_*m_{\rm h}^{-2/3}.
    \label{eq:beta}
\end{align}
After the disruption, roughly half of the stellar debris falls back at the pericenter and emits luminous X-ray flares \textcolor{black}{(see, e.g., \citealt{Rees:88aa} and \citealt{Evans:89aa})}. Then, the debris circularize and form an accretion disc \textcolor{black}{(see, e.g., \citealt{Bonnerot:20aa})}. To date, around $\sim 20$ robust X-ray TDEs have been observed (see the review by \citealt{Saxton:20} and references therein). Another $\sim 30$ TDEs have been detected in the optical (see the review by \citealt{vanvelzen:2020} and references therein). Few of them present both X-ray and optical radiation, but the majority seem to be part of a different electromagnetic class of TDEs. The mechanism behind this kind of emission is still theoretically debated (see, e.g., \citealt{Lodato:11aa}, \citealt{roth:20aa}, \citealt{Bonnerot:20aa}). Thanks to future surveys and a growing number of electromagnetically detected TDEs it will be possible to collect more information and solve this problem.\\
\indent TDEs also emit GWs. In particular we can distinguish between i) GW burst emitted when the star is torn apart at $r_{\rm p}$ (\textcolor{black}{variation of the mass quadrupole of the BH-star system;} see, e.g., \citealt{Kobayashi_2004}), ii) GW production when the star is stretched and compressed by BH tides while approaching the pericenter (\textcolor{black}{variation of the internal mass quadrupole of the star;} see, e.g., \citealt{Guillochon:09aa} and \citealt{Stone:13aa}) and iii) GW radiation at later stages (see, e.g., \citealt{Kiuchi:11aa} and \citealt{Toscani:19aa} for emission after the circularization). Focusing on i), that for standard values of the parameters involved in the disruption is the strongest contribute to the GW emission, we have that the GW amplitude, $h$, and the duration of the signal, $\tau$, can be estimated as (see, e.g., \citealt{Kobayashi_2004}, \citealt{Toscani:20aa})
\begin{align}
 h &\approx \beta \times \frac{m_{*}^{4/3}m_{\rm h}^{2/3}}{d_{20}r_{*}}\times 2\times 10^{-26},\label{eq:amplitude}\\
 \tau & \approx\beta^{-3/2}\times m_{*}^{-1/2}r_{*}^{3/2}\times 10^4 \,\text{s}\label{eq:duration},
\end{align}
where $d_{20}$ is the distance in units of 20 Mpc. 
 \begin{figure*}
    \centering
    \includegraphics[width=0.246\textwidth]{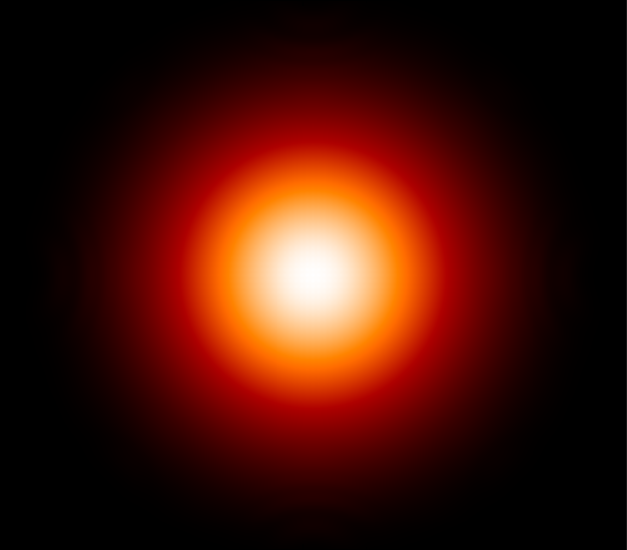}
\includegraphics[width=0.246\textwidth]{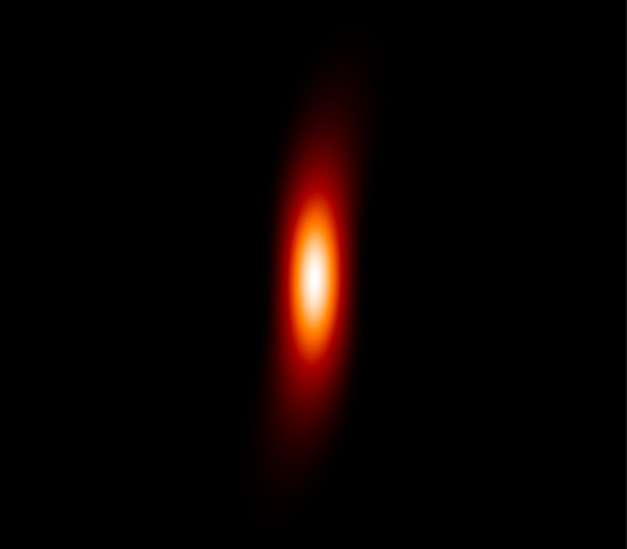}
\includegraphics[width=0.246\textwidth]{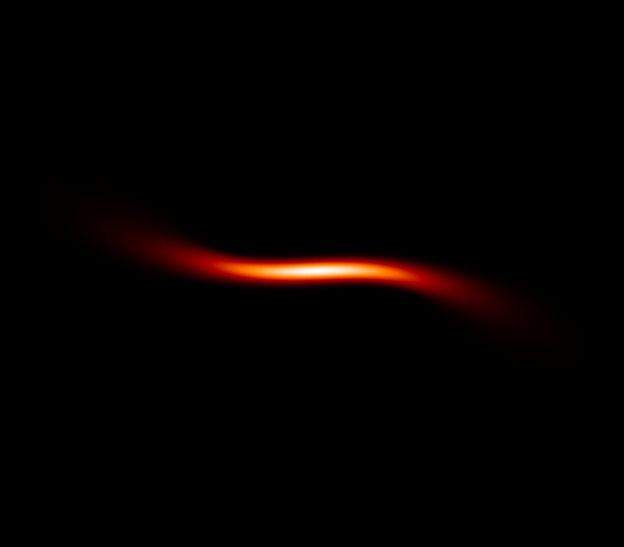}
\includegraphics[width=0.246\textwidth]{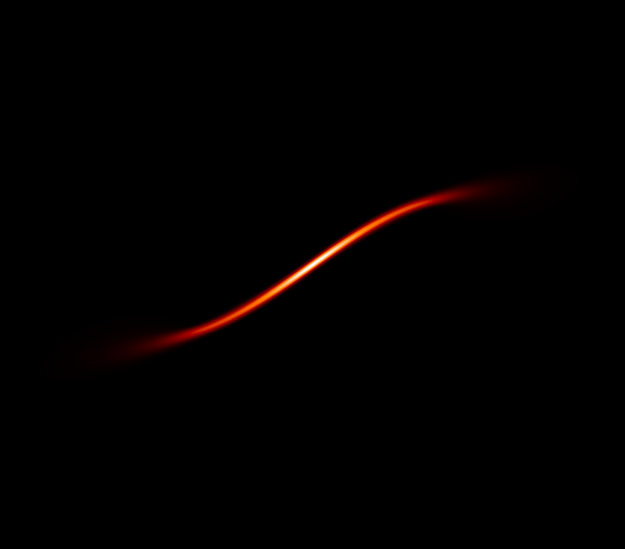}
\caption{\textcolor{black}{Surface density plots for the TDE simulation described in Section \ref{standard}. From left to right: star at the beginning of the simulation (first panel), during the disruption (second panel), leaving the pericenter (third and fourth panel). These snapshots are generated with SPLASH \citep{Price:07aa}, an SPH visualization code.}} 
    \label{fig:snapshot}
\end{figure*}
\subsection{Tidal disruption events with \textsc{GR-PHANTOM}}
\label{standard}

\begin{figure*}
    \centering
    \includegraphics[width=0.8\textwidth]{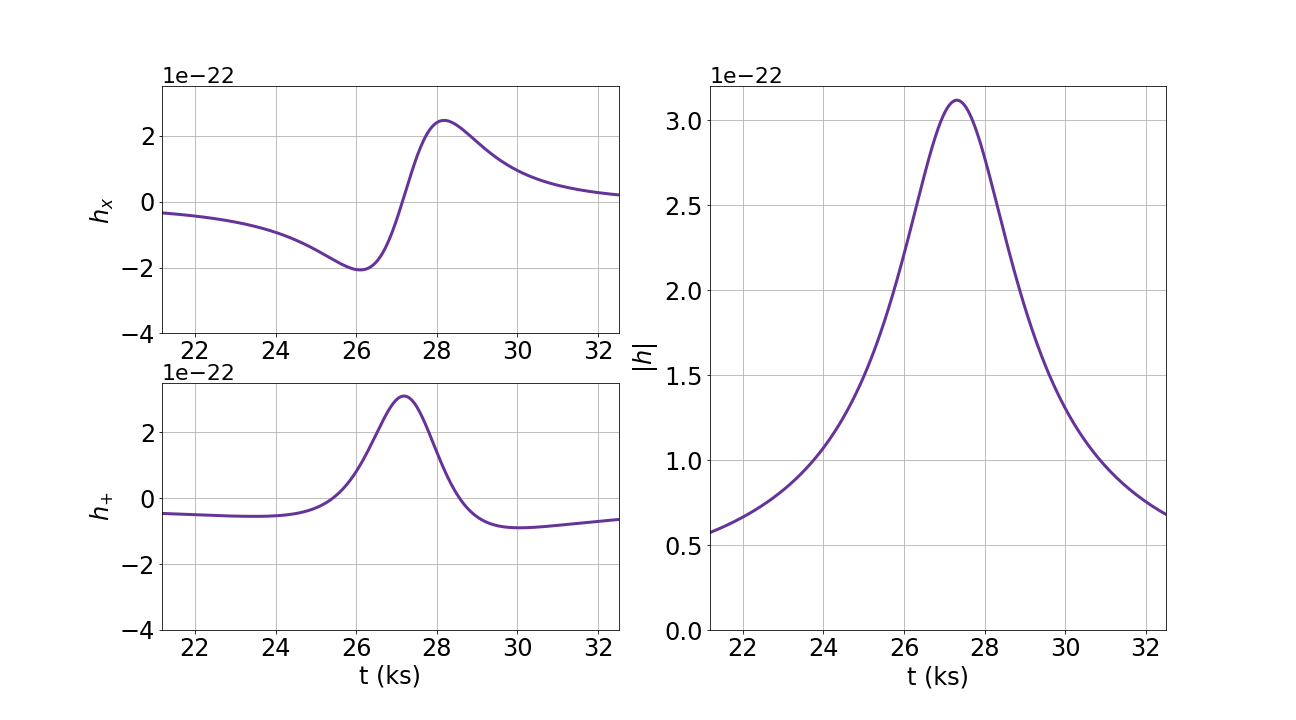}
\caption{GW emission from a TDE of a MS star with $M_*=1\text{M}_{\odot}$ and $R_*=1\text{R}_{\odot}$ on a parabolic orbit around a Schwarzschild BH with $M_{\rm h}=10^{6}\text{M}_{\odot}$, plotted with respect to time (in kiloseconds). The penetration factor is $\beta=1$ and the source is at $d=20 \text{Mpc}$. \textbf{Top left}: $\times$ polarization seen face-on. \textbf{Bottom left}: $+$ polarization seen face-on. \textbf{Right}: \textcolor{black}{Root-square-sum} amplitude.} 
    \label{fig:tde_gr}
\end{figure*}
\begin{figure}
    \centering
    \includegraphics[width=0.4\textwidth]{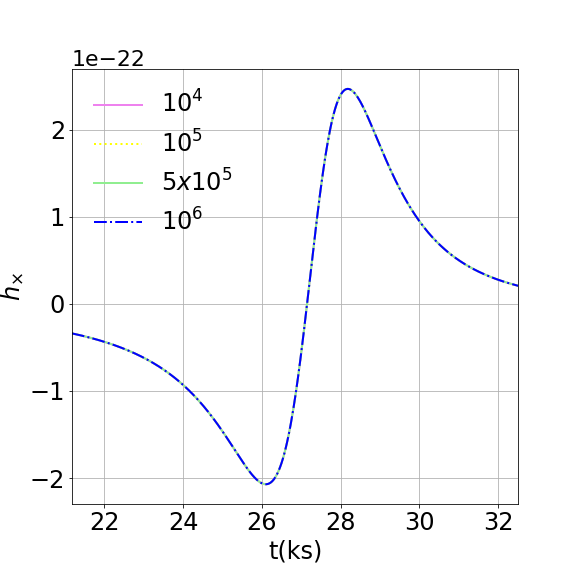}
\caption{GW signal for a TDE of a Sun-like star, as described in figure \ref{fig:tde_gr}. We plot the $\times$ polarization with respect to the time (in kiloseconds) for $10^4$, $10^5$, $5\times 10^5$ and $10^6$ particles. All the curves obtained coincide.} 
    \label{fig:conv_strain}
\end{figure}
We first simulate the TDE of a main sequence (MS) star of $1\text{M}_{\sun}$ and $1\text{R}_{\sun}$ on a parabolic orbit around a non-rotating BH $M_{\rm h}=10^{6}\text{M}_{\sun}$, with $\beta=1$, in order to compare our results with the same case presented in \citealt{Kobayashi_2004}. We assume a source located at a distance of $20\,\text{Mpc}$ and we simulate a polytropic $\gamma=5/3$ sphere with $10^5$ equal-mass particles. Some snapshots of this simulation are illustarted in Figure \ref{fig:snapshot}.\\
\indent {The GW signal is plotted in Figure \ref{fig:tde_gr}, with respect to time (in kiloseconds), where we show the waveforms $h_{+}(t), h_{\times}(t)$ and also the \textcolor{black}{GW strain approximated as} \textcolor{black}{root-square-sum} amplitude $|h(t)|=\sqrt{ |h_{+}(t)|^{2}+|h_{\times}(t)|^{2}}$. First, we note that the maximum value of the signal is $\approx 3\times 10^{-22}$, while the duration $\approx 10 \,\text{ks}$, both values being consistent with equations \eqref{eq:amplitude} and \eqref{eq:duration} respectively}. These results are also consistent with the plot in the top panel of Figure 5 by \citealt{Kobayashi_2004}.\footnote{{There is a minus sign of difference between our plot of $h_{+}$ and the one presented in \citealt{Kobayashi_2004}, probably due to a $\pi$ difference in the viewing angle.}}


\subsubsection{Resolution check}
We perform a resolution test to see how many particles we need to reproduce the signal illustrated in Figure \ref{fig:tde_gr}. Thus, we repeat the same simulation with $10^4$, $10^5$, $5\times 10^5$ and $10^6$ particles. As shown in figure \ref{fig:conv_strain}, $10^4$ particles are already enough to obtain the desired curve, indicating that GWs are only sensitive to the bulk mass distribution. \textcolor{black}{In addition to this, we have also calculated the relative differences between the curves, to be sure about the convergence. We see that the absolute errors are of order $\sim 10^{-4}$, thus confirming the convergence.}

\subsection{TDE gravitational catalogue}
In this Section we present the simulation included in our atlas and the parameter space explored. The catalogue is available online at \url{https://gwcataloguetdes.fisica.unimi.it}.\\

\noindent \textbf{Orbital inclination.} In the GW catalogue we assume that the stellar orbit lies in the $\pmb{x}-\pmb{y}$ plane (see geometry in Figure \ref{fig:angles}). In practice, in \textsc{grphantom}, in the case of a spinning BH, the coordinate choice naturally has the $\pmb{z}$ axis pointing in the BH spin direction. In these cases, we then simulate an encounter with an inclined stellar orbit with respect to the $\pmb{z}$ axis, by the desired angle $\theta$, and then rotate the results back into the $\pmb{x}-\pmb{y}$ plane \textit{before computing} the GW emission.\\

\noindent \textbf{TDE of a Sun-like star.} {We consider a Sun-like star on an orbit with three possible inclination angles with respect to the BH spin: $\theta=0-30^{\circ}-60^{\circ}$. For each value of this angle, we span the BH mass over the following values: $m_{\rm h}=10^5-10^6-10^7$. Each BH has three possible values of the spin: $a=0-0.5-0.9$. For every scenario we consider both elliptical orbits (with eccentricities $e$=0.6, $e$=0.8) and parabolic ($e$=1). All orbits are prograde. For $m_{\rm h}=10^{7}$ we limit to $\beta=1$ (because for higher penetrations the star is swallowed whole before disruption), while for the other BH masses we consider three values of $\beta$: 1-2-5. Additionally, for $m_{h}=10^{6}$ we have decided to investigate also the retrogade scenario ($\theta=180$) with values of the $\beta$ parameters equal equal to 1 and 2.}

{Overall, we have performed 207 simulations. If we remove those with spinless BH and different $\theta$, that are equivalent to each other thanks to spherical symmetry, the effective number of simulations is 147.}\\
 
\noindent \textbf{TDE of a MS star.} We consider a MS star with $M_{*}=10\,\text{M}_{\odot}$, with a radius in solar units (see \citealt{Kippenhan:90aa})
\begin{equation}
    r_{*}= m_{*}^{0.57}\approx 3.7.
\end{equation}
We assume the star to be disrupted by a $10^{7}\,\text{M}_{\odot}$ BH, with spin $a=0-0.5-0.9$. We examine three values of the eccentricities $e=0.6-0.8-1$ and we keep the penetration factor fixed to 1. All the orbits are prograde. The inclination angles are $\theta=0-30^{\circ}-60^{\circ}$. Overall, we have performed 27 simulations. Due to the spherical symmetry, the effective number of simulations is 21.\\

\noindent \textbf{TDE of a WD.} We consider a WD star of mass $0.5\,\text{M}_{\odot}$ and radius $10^{-2}R_{\odot}$ (see \citealt{Shapiro:83aa}) disrupted by a $10^{4}\,\text{M}_{\odot}$ BH, with spin $a=0-0.5-0.9$. We investigate three values of the eccentricities $e=0.6-0.8-1$ and we take $\theta=0$. The penetration factor is fixed to 1. {Since the stellar mass is far from the Chandrasekar limit, we can simulate the star as a polytrope with $\gamma=5/3$ instead of $\gamma=4/3$}. Overall, we have performed 9 simulations.\\

At the time of writing, our online catalogue contains a total number of waveforms equals to 243. However, we consider this a \textit{live} catalogue and our plan is to continue to expand it over time, in order to cover more parameter space and make it as comprehensive as possible.\\

\noindent \textbf{Distance and orientation of the events} {In our catalogue, the star is always assumed to lie at a distance of 1 Mpc.} The user can easily rescale the results to any assumed distance. Also, for each simulation the catalogue includes the GW emission computed assuming the following values of the angles $(\eta, \phi)$: $(0,0)$,$(30,0)$,$(60,0)$,$(90,0)$. However, we also offer the components of the second time derivative of the moment of inertia, so the user can compute easily the GW signal with any geometry, by using equation \eqref{eq:strain}.

\section{Discussion}
\label{sec:5}
Here we discuss some of the results from our simulations. The distance of the source is fixed at $20\,\text{Mpc}$ and the polytropic index is $\gamma=5/3$. All the calculations are performed in Kerr metric expressed with Boyer-Lindquist coordinates. The transformations from code to physical units are given by
\begin{align}
    t_{\rm phys}&=4.96\,\text{s}\times t_{\rm code}\times\left(\frac{M_{\rm h}}{10^6\,\text{M}_{\odot}}\right),\\
    l_{\rm phys}&=0.0098\,\text{au}\times l_{\rm code}\times\left(\frac{M_{\rm h}}{10^6\,\text{M}_{\odot}}\right),\\
    M_{\rm phys}&=M_{\rm code}\times M_{\rm h}\,,
\end{align}
\textcolor{black}{where $t$, $l$ and $M$ stand for time, length and mass respectively.}
\subsection{Changing the physical parameters}
\subsubsection{Spin}
\label{spin}
\begin{figure*}
    \centering
    \includegraphics[width=0.8\textwidth]{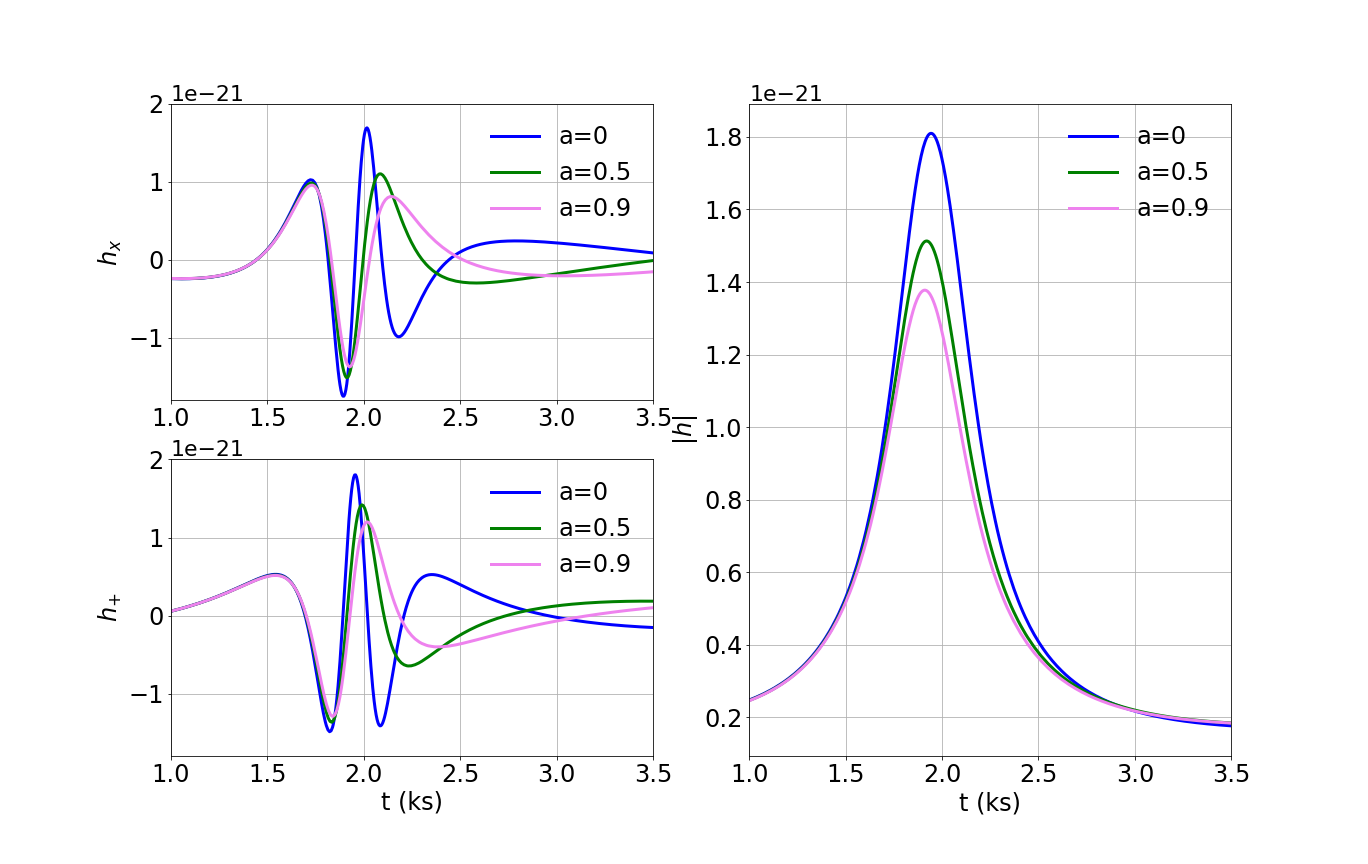}
\caption{Sun-like star disrupted by a $10^{6}\,\text{M}_{\odot}$ BH. The orbital inclination is $\theta=0$, the penetration factor is $\beta=5$ and the eccentricity $e=0.6$. Prograde orbits. \textbf{Top left}: $\times$ polarization seen face-on. \textbf{Bottom left}: $+$ polarization seen face-on. \textbf{Right}: Related \textcolor{black}{root-square-sum} amplitude. All the curves are plotted with respect to time (in kiloseconds). The three colours stand for three different BH spin values: blue $a=0$, green $a=0.5$ and pink $a=0.9$. \textcolor{black}{We can see that increasing the spin changes the shapes of the polarization amplitudes during and after the pericenter, shifting the peak of the waveforms at later times and lowering it, and that the peak of the root-sum-square amplitude decreases while the spin increases. The peak for $a=0.9$ is a factor 1.4 smaller than the peak for spinless BH.}} 
    \label{fig:spin_changes}
\end{figure*}
\begin{figure}
\centering
    \includegraphics[width=0.4\textwidth]{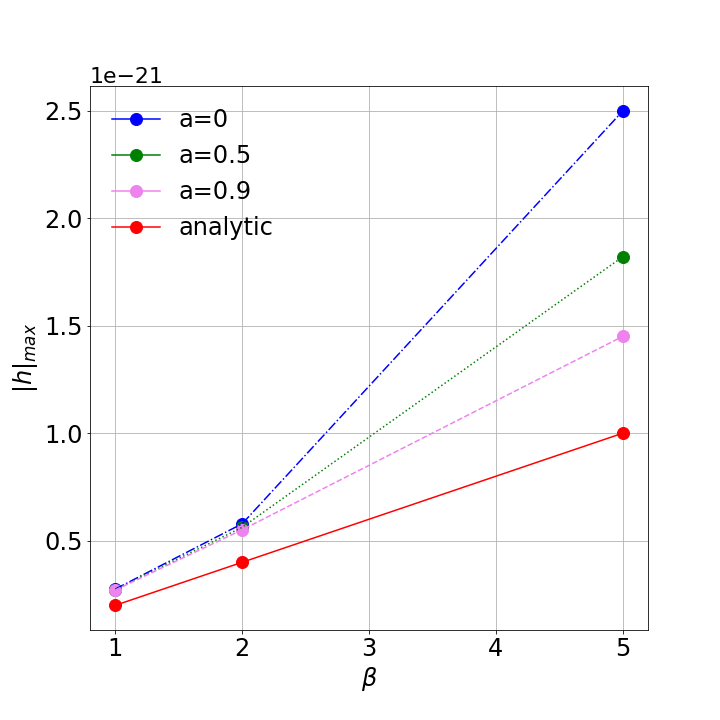}
\caption{Maximum \textcolor{black}{root-square-sum} amplitude versus $\beta$ for a Sun-like star disrupted by a $10^{6}\text{M}_{\odot}$ BH. The orbital inclination is $\theta=0$, the eccentricity $e=0.6$, prograde orbits. The three different colours stand for three different BH spins: blue $a=0$, green $a=0.5$ and pink $a=0.9$. \textcolor{black}{The red solid line shows the linear relationship between the penetration factor and the maximum GW amplitude predicted from Equation \eqref{eq:amplitude}. This line is lower than the others since the GW amplitude increases for higher values of the eccentricity, while Equation \eqref{eq:amplitude}  refers to null eccentricity (cf. paragraph \ref{sec:ecc}).}}
\label{fig:betaspin}
\end{figure}
\begin{figure}
\includegraphics[width=0.4\textwidth]{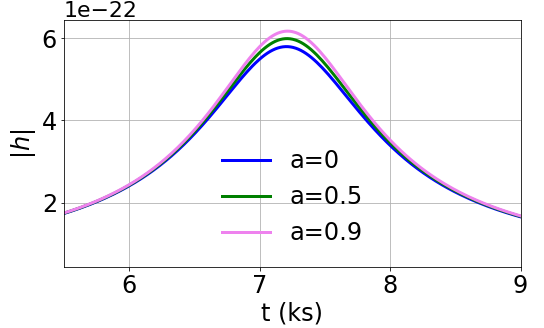}
\caption{\textcolor{black}{Root-square-sum} amplitude plotted with respect to time (in kiloseconds). We consider a Sun-like star disrupted by a $10^{6}\text{M}_{\odot}$ BH; the penetration factor is $\beta=2$, the eccentricity of the stellar orbit is $e=0.6$ and the orbit is retrograde ($\theta=180^{\circ}$). The three different colors refer to three different spins: blue $a=0$, green $a=0.5$ and pink $a=0.9$. \textcolor{black}{The difference between the a = 0.9 and a = 0 curves is roughly $6\%$.}}
\label{fig:retro}
\end{figure}
\begin{figure*}
    \centering
    \includegraphics[width=0.7\textwidth]{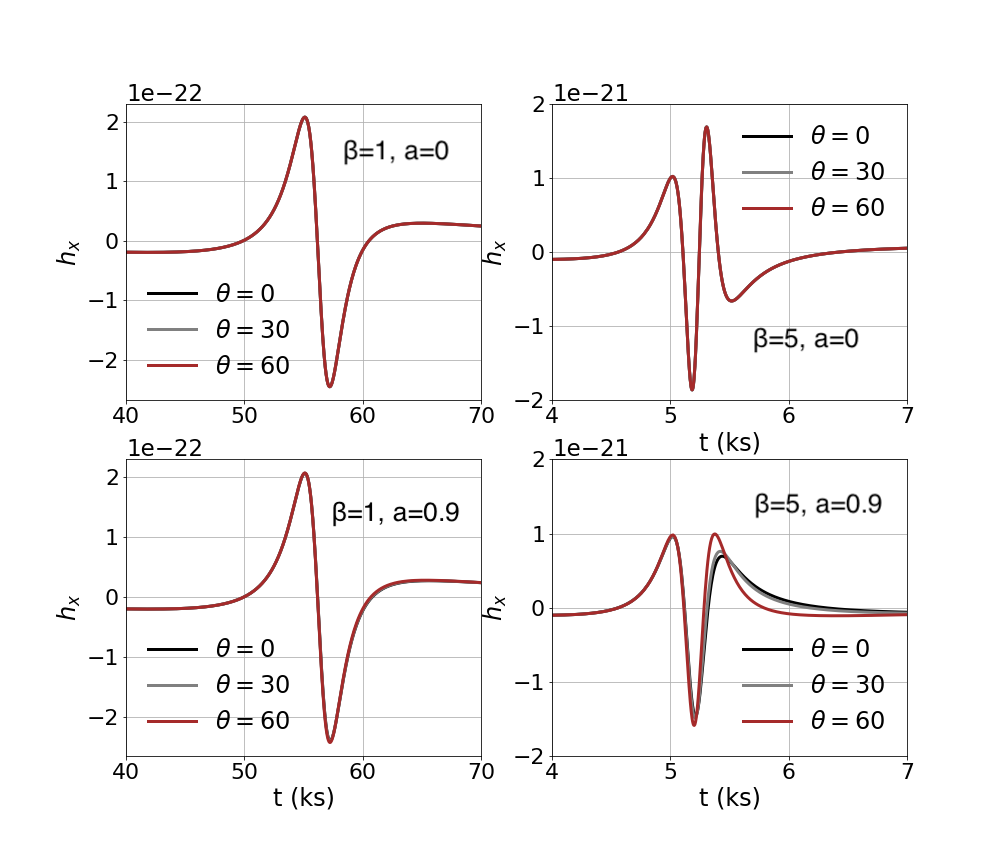}
\caption{\textcolor{black}{Sun-like star disrupted by a $M_{\rm h}=10^{6}\,\text{M}_{\odot}$ BH. Eccentricity e=0.8, prograde orbits. \textbf{Top left}: $\times$ polarization seen face-on. $\beta=1$, spinless BH. \textbf{Bottom left}: $\times$ polarization seen face-on. $\beta=1$, $a=0.9$. \textbf{Top right}: $\times$ polarization seen face-on. $\beta=5$, $a=0$. \textbf{Bottom right}: $\times$ polarization seen face-on. $\beta=5$, $a=0.9$. All the curves are plotted with respect to time (in kiloseconds). The black curve stands for $\theta=0$, the grey for $\theta=30^{\circ}$ and the brown for $\theta=60^{\circ}$. From this Figure, it is possible to see that different inclination angles of the stellar orbits affect the GW emission only when we have both high values of $\beta$ and high BH spins.}} 
    \label{fig:theta_changes}
\end{figure*}
\begin{figure}
    \centering
    \includegraphics[width=0.4\textwidth]{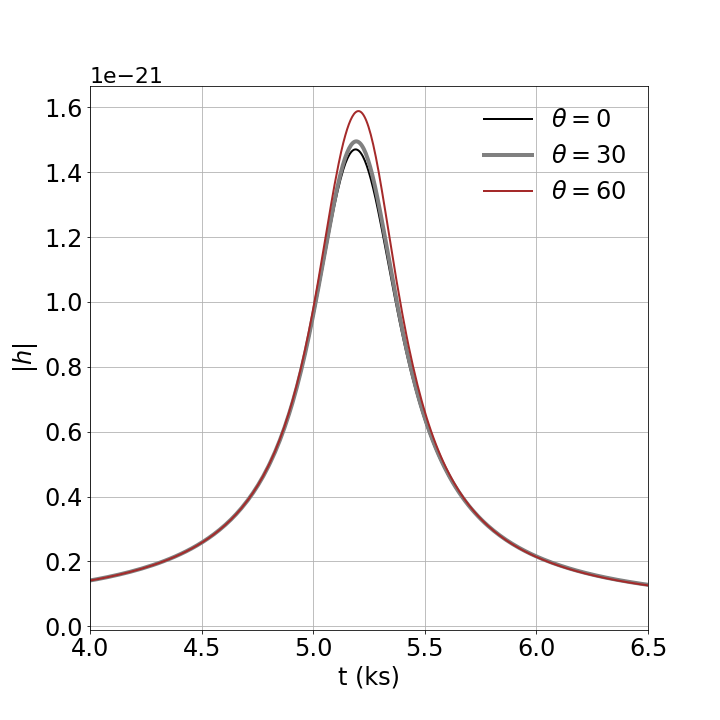}
    \caption{\textcolor{black}{Root-square-sum} signal plotted with respect to time (in kiloseconds) for the same scenario as the one illustrated in the right panel of Figure \ref{fig:theta_changes}. \textcolor{black}{The peak decreases together with the inclination angles ( $\approx 7\%$).}}
    \label{fig:peaktheta}
\end{figure}
First, we have investigated if the spin affects the GW signal, considering prograde orbits with respect to the spin of the BH. We have found that, if compared to the static BH scenario,  i) the spin changes the shapes of the polarization amplitudes during and after the passage at the pericenter, {shifting the peak of the waveforms at later times and lowering it}, ii) the peak of the \textcolor{black}{root-square-sum} decreases while the spin increases.\\
\indent In Figure \ref{fig:spin_changes}, we illustrate a close orbit in the $\pmb{x}-\pmb{y}$ plane with eccentricity $e=0.6$ and penetration factor $\beta=5$. We plot the polarizations $\times$ (top left) and $+$ (bottom left) seen face-on and the related \textcolor{black}{root-square-sum} amplitude $|h|$, versus time (in kiloseconds). In each panel we consider three values for the spin: $a=0$ (blue), $a=0.5$ (green) and $a=0.9$ (pink). Deviations from the static BH case are clearly visible in each plot, starting from $\approx 1.8$ ks, that corresponds to the star passing through $r_{\rm p}$. The peak in the strain for $a=0.9$ is smaller than the peak for spinless BH by a factor $\approx 1.4$. \\
\indent In Figure \ref{fig:betaspin}, we plot $|h|_{\rm max}$ for the same scenario as before but as a function of $\beta$. We can see that $h_{\rm max}$ starts to be affected by the spin already at low $\beta$ ($\approx 2$), even though the spin effect can be appreciated more easily at $\beta \approx 5$. Thus, the proportionality between the maximum value of the GW signal \textcolor{black}{and} $\beta$ predicted from theory (cf. \ref{eq:amplitude}), seems to hold. For the case of rotating BHs the proportionality constant is $\approx 1$ as expected, while for the spinless case this constant seems to be between $1 \sim 2$.\\
\indent {Contrary, for retrograde orbits ($\theta=180^{\circ}$) we see the GW signal increases for higher values of $a$. In Figure \ref{fig:retro}, we plot $|h|$ versus time (in kiloseconds), for a Sun-like star disrupted by a $10^{6}\text{M}_{\odot}$ BH. The penetration factor is equal 2 and the eccentricity is 0.6. We consider three different values of the BH spin: $a=0$ (blue), $a=0.5$ (green) and $a=0.9$ (pink). The difference between the peak corresponding to $a=0.9$ BH respect to the spinless BH is roughly $6\%$.} From our results, it seems that GW amplitudes are stronger for high spin retrograde orbits and weaker for high spin prograde orbits. \textcolor{black}{We interpret this result as due to the higher angular frequency that a retrograde orbiter experiences. In fact, the orbital frequency of a test particle orbiting around a spinning BH is \citep{Bardeen:72aa}
\begin{align}
    |\nu_{\phi}|=\frac{c^3}{2\pi GM_{\rm h}}\left| \frac{1}{(2r/r_{\rm g})^{3/2}\pm a}\right|.
\end{align}
From the above equation we see that the angular frequency increases for higher retrograde (-) BH spins.}


\subsection{Changing the orbital parameters}
\subsubsection{Inclination angle}
If the central BH is not rotating, the GWs should be independent of the orbital inclination $\theta$, thanks to the spherical symmetry of the system. The results of our simulations proved this assumption i) both in the case of spinless BH and ii) for star far enough from the BH so that the effects of rotation are negligible. Contrarily, when the the spin \textit{does} modify the shape of the signal, the condition of spherical symmetry does not hold anymore and thus different values of $\theta$ change the signal emitted by the TDE.\\
\indent In Figure \ref{fig:theta_changes}, we illustrate the GW polarization amplitudes for a close orbit ($e=0.8$) of a Sun-like star disrupted by a $10^{6}\,\text{M}_{\odot}$ BH. We plot the $\times$ polarization seen face-on for $\beta=1$ with spinless BH (top left) and spin $a=0.9$ (bottom left) and for \textcolor{black}{$\beta=5$ with spinless BH (top right)} and spin $a=0.9$ (bottom right). The black curve refers to $\theta=0$, the grey for $\theta=30^{\circ}$ and the brown for $\theta=60^{\circ}$. We see that the values of $\theta$ have an effect on the amplitudes only in the panel on the \textcolor{black}{lower} right, where the changes are appreciable during and following the passage at the pericenter (cf. paragraph \ref{spin}). As for the strain, the peak slightly decreases together with the inclination angles ($\approx 7\%$), as illustrated in Figure \ref{fig:peaktheta}.

\subsubsection{Eccentricity}
\label{sec:ecc}
\begin{figure*}
    \centering
    \includegraphics[width=0.8\textwidth]{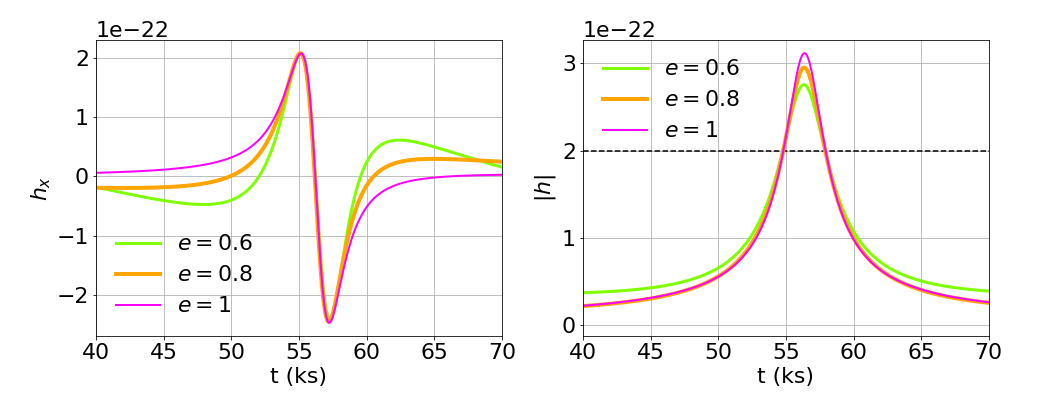}
\caption{Sun-like star disrupted by a $M_{\rm h}=10^{6}\,\text{M}_{\odot}$ spinless SMBH. The orbital inclination is $\theta=0$, the penetration factor is $\beta=1$.  \textbf{Left}: $\times$ polarization seen face-on. \textbf{Right}: related strain. All the curves are plotted with respect to time (in kiloseconds). The light green curve refers to $e=0.6$, the orange to $e=0.8$ and the magenta to $e=1$. The black horizontal dashed line is the analytical \textcolor{black}{estimate} of the strain (see equation \ref{eq:amplitude}).} 
    \label{fig:ecc_change}
\end{figure*}
An additional orbital parameter that can have an impact on the GW signal emitted by tidally disrupted stars is the geometry of the orbit. We consider three values of the eccentricity: $e=0.6$ and $e=0.8$ for elliptical orbits and $e=1$ for parabolic orbits. From the numerical outputs of our simulations, we find that different values of the eccentricities modify the polarization amplitudes both \textit{before} and \textit{after} the passage at the pericenter. As for the strain, the peak increases with the eccentricity. This can be explained as followed. The GW burst from TDEs is produced around the pericenter, when the star is torn apart by the BH tides. At the pericenter, assuming Keplerian velocity, $v_{\rm kepl}$ reads
\begin{align}
   v_{\rm kepl}=\left[ \frac{GM_{\rm h}}{r_{\rm p}}(1+e)\right]^{1/2}\propto (1+e)^{1/2}.
\end{align}
We see that for higher eccentricities the star goes through the pericenter faster. This results in a stronger gravitational signal.\\
\indent In Figure \ref{fig:ecc_change}, we illustrate the scenario of a Sun-like star disrupted by a $10^{6}\,\text{M}_{\odot}$ spinless BH, with $\beta=1$. We can see that the differences in the polarization amplitudes for each eccentricity are well visible. As for the strain, the black dashed line represent the analytical estimates, that is a factor 1.5 lower than the peak for the parabolic trajectory ($(1+e)^{1/2}=2^{1/2}\approx 1.4$) consistent with our estimates of Equation \eqref{eq:amplitude}, since the analytical estimates assumes a circular orbit ($e \approx 0$).
\begin{figure}
    \centering
    \includegraphics[width=0.4\textwidth]{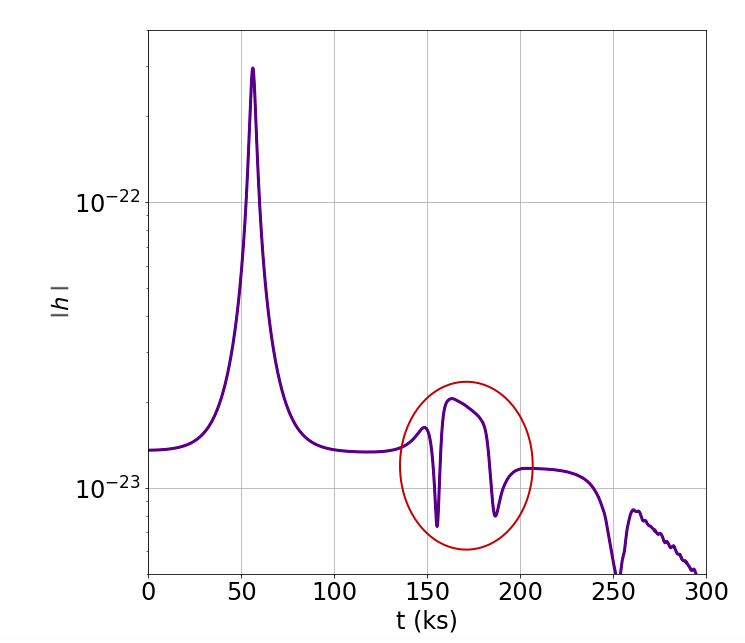}
    \caption{GW signal produced by a Sun-like star disrupted by a Schwarzschild BH with mass $M_{\rm h}=10^{6}\,\text{M}_{\odot}$ BH. The orbital inclination is $\theta=0$, the penetration factor is \textcolor{black}{$\beta=1$}, the eccentricity $e=0.8$. During the first passage at the pericenter ($\approx 70$ ks) the star is torn apart by the BH tides and the first GW burst is produced. Then the debris starts the circularization process around the BH. Here we consider $\sim 2$ returns to the pericenter. With the red circle we outline the bump in the strain due to the first return $r_{\rm p}$.}
    \label{fig:5_orbs}
\end{figure}

\subsection{Return to pericenter}
We have investigated if the debris, falling back to the pericenter, produce a GW signal comparable to the one emitted when the star is torn apart. We consider a Sun-like star disrupted by a static $10^{6}\,\text{M}_{\odot}$, on a close orbit ($e=0.8$) in the $\pmb{x}-\pmb{y}$ plane with \textcolor{black}{$\beta=1$}. We evolve the simulation for $\sim 3$ orbits. From Figure \ref{fig:5_orbs}, we can see that, when the debris falls back to $r_{\rm p}$ for the first time ($t\approx 170$ ks), they produce a bump in the strain, that is a factor $\approx 10$ smaller than the disruption signal. For the following passages at the pericenter, the gravitational contribution is even lower (see \citealt{Toscani:19aa} for further discussion on this secondary emission).\\
\indent It would be interesting if this second bump could be seen by future interferometers. Indeed, when we talk about the multimessenger prospects for TDEs, we think to detect the GW signal from the disruption followed after some days by the electromagnetic counterpart. But, if this second bump could be revealed, the electromagnetic signal would be accompanied by GW detection and thus we would be able to collect more and complementary information on the circularization process and possibly on the emission mechanism that fuel TDE flares. 

\subsubsection{Energy lost by GW emission}
In the previous example we should have considered the energy lost by the system via GWs, $E_{\rm gw}$, to be sure that the orbit done by the debris is the correct one. \textcolor{black}{Yet, we expect $E_{\rm gw}$ to be small enough relative to the orbital energy of the star and so negligible.} Here we present a quick way to estimate this quantity.\\
\indent \textcolor{black}{If we consider a Sun-like star disrupted by a $10^6\text{M}_{\odot}$ BH with $\beta=1$, the third time derivative of the moment of inertia of the system can be approximated as
\begin{align}
   \dddot{M}\approx \frac{M_{*}r_{\rm t}^2}{\tau^3}.
\end{align}
In the above Equation we have assumed the star as a point-mass particle orbiting around the BH at a distance of $r_{\rm t}$. Thus, we have calculated the moment of inertia of the system as $Mr_{\rm t}^2$ and we have derived its third derivative with respect to time dividing it by the cube of the period, 
$\tau^3$. Consequently, the luminosity associated with the GW burst is}
\begin{align}
    L_{\rm gw}\approx \frac{G}{5c^5}\left[M_{*}r_{\rm t}^{2}\left(\frac{1}{\tau}\right)^{3}\right]^2.
\end{align}
As a result, the energy lost by the system over a period will be approximately $E_{\rm gw}\approx 10^{37}\,\text{erg}$. The absolute value of the orbital energy for the same system reads
\begin{align}
    E_{\rm orb}=\frac{GM_{\rm h}M_{*}}{2r_{\rm t}}\approx 10^{52}\,\text{erg}.
\end{align}
Thus we see that $E_{\rm gw}$ is \textcolor{black}{fifteen} orders of magnitude lower than the orbital energy of the system and, consequently, it is safe to neglect the back-reaction of the GW emission.

\subsection{Angular distribution of quadrupole radiation}
\begin{figure*}
    \centering
    \includegraphics[width=0.8\textwidth]{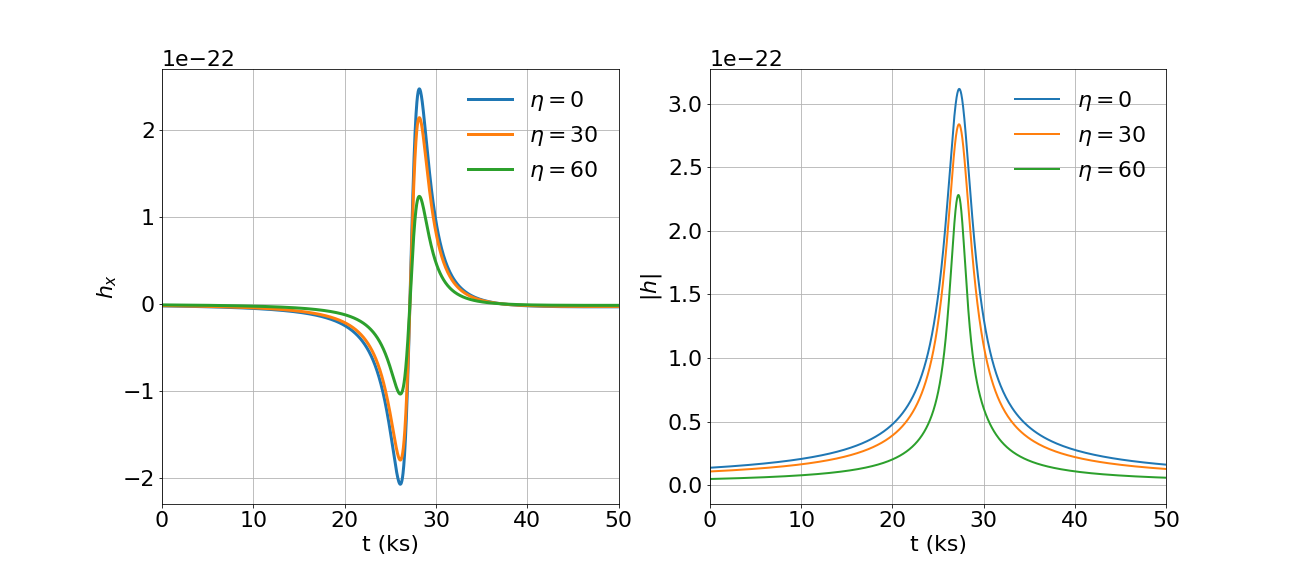}
\caption{Sun-like star disrupted by a $10^{6}\,\text{M}_{\odot}$ spinless SMBH. The orbital inclination is zero, the penetration factor is $\beta=1$.  \textbf{Left}: $\times$ polarization. \textbf{Right}: related strain. All the curves are plotted with respect to time (in kiloseconds). The blue curve refers to $\eta=0$, the orange to $\eta=30^{\circ}$ and the green to $\eta=60^{\circ}$.} 
    \label{fig:eta_change}
\end{figure*}
\begin{figure}
    \centering
    \includegraphics[width=0.35\textwidth]{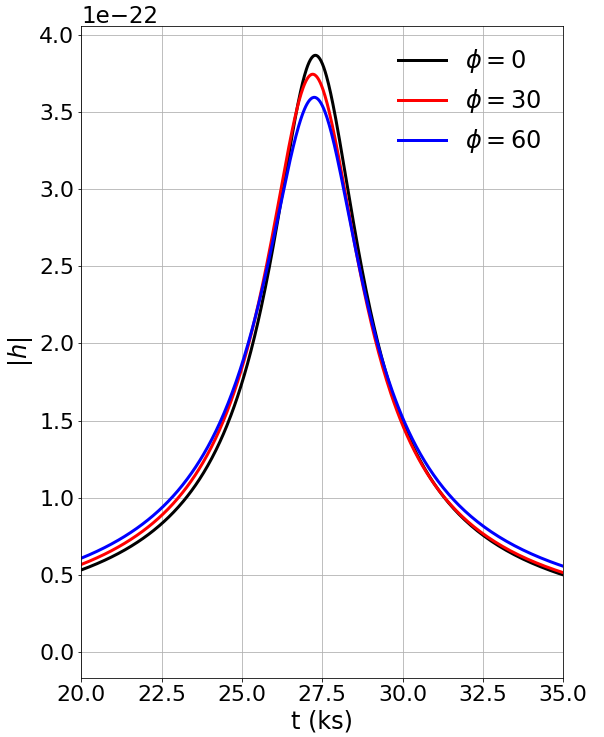}
    \caption{Sun-like star disrupted by a $10^{6}\,\text{M}_{\odot}$ spinless SMBH. We have $\eta=30^{\circ}$, while the penetration factor is $\beta=1$.  \textcolor{black}{Root-square-sum} amplitude plotted with respect to time (in kiloseconds). The black curve is for $\phi=0$, while the bue curve for $\phi=60^{\circ}$. All the curves are plotted with respect to time (in kiloseconds).}
    \label{fig:phichange}
\end{figure}
We recall that $\eta$ is the  angle between the los and the normal to the orbit, while $\phi$ is the angle between the projection of the line of sight on the $\pmb{x}-\pmb{y}$ plane and the pericenter.\footnote{So far we have kept this angle fixed to 0, in the same way we present the waveforms in the catalogue.}

\subsubsection{Changing $\eta$}
We have investigated what happens if we consider $\phi=0$ and we change $\eta$. Figure \ref{fig:eta_change} shows the $\times$ polarization (left) and the strain (right) with respect to the time (in kiloseconds). We consider the observer along the normal to the orbit (blue curve), inclined by $30^{\circ}$ (orange curve) and by $60^{\circ}$ (green curve) respect to the $\pmb{z}$ axis. The strain is strongest when the observer see the signal face-on. \textcolor{black}{As we could expect, the signal is strongest when it propagates along the direction of the line of sight, otherwise it decreases.}

\subsubsection{Changing $\phi$}
We have investigated what happens if we consider $\eta=30^{\circ}$ and we change $\phi$. {We can see from Figure \ref{fig:phichange} that if the projection of the los moves away from the pericenter sight, the signal decreases.}

\section{Conclusions}
\label{sec:6}
In this paper, we have implemented a new method for the derivation of GWs in \textsc{grphantom}, an SPH code including GR dynamics. Our tool is based on the discretization of the second time derivative of the moment of inertia and then the GWs are calculated in the weak field approximation. We test it for some well-known systems (isolated star, binary far from the merger phase). Then, we use it to derive the GW emission from TDEs, spanning the space of parameters. The main results we have obtained from the simulations are the following:
\begin{enumerate}
    \item the spin affects the GW polarization amplitudes during and after the passage at the pericenter. The strain decreases with the increasing of the spin for prograde orbits, while the opposite occurs for retrograde orbits;
    \item the orbital inclination affects the GW emission from TDEs only if the BH is spinning and the star is close enough to feel the effect of the spin;
    \item different eccentricities modify the polarization amplitudes. The strain increases for higher eccentricities, due to the fact that the velocity at pericenter is $\propto (1+e)^{1/2}$;
    \item after the disruption, the debris falling back at the pericenter produces a bump in the strain $\sim 1$ order of magnitude lower that the signal at the pericenter;
    \item the GW emission is stronger if we observe the system face on and if the signal is along the axis of the pericenter.
\end{enumerate}
\textcolor{black}{From these results, it seems likely that TDEs discovered through GWs will be biased towards face-on low-spin prograde TDEs/high-spin retrograde TDEs, with very eccentric orbits.}\\
\indent All the {results of the simulations} can be found online at the following \url{https://gwcataloguetdes.fisica.unimi.it}. We believe that such a catalogue can be a useful resource of templates to easily classify the GW emission from TDEs once this kind of signal will be able to be detected by future interferometers.

\section*{Acknowledgments}
This project has received funding from the European Union's Horizon 2020 research and innovation program under the Marie Sk\l{}odowska-Curie grant agreement NO 823823 (RISE DUSTBUSTERS project). We acknowledge CPU time on OzSTAR, funded by Swinburne University and the Australian Government. 

\section*{Data Availability Statement}
The data underlying this article are available at \url{https://gwcataloguetdes.fisica.unimi.it}, and can be accessed freely by any user.

\bsp	
\label{lastpage}
\end{document}